\documentclass[12pt]{iopart}

\usepackage{iopams}
\usepackage {graphicx}
\begin{document}

\title{Description of hysteretic current-voltage characteristics of SNS
junctions}

\author{D M Gokhfeld}

\address{L.V. Kirensky Institute of Physics SD RAS,
Krasnoyarsk, 660036, Russia}
\ead{gokhfeld@iph.krasn.ru}
\begin{abstract}
Simplified model for current-voltage characteristics of weak links
is suggested. It is based on an approach considering the multiple
Andreev reflection in metallic Josephson junction. The model allows
to calculate current-voltage characteristics of the superconductor -
normal metal - superconductor junctions with different thicknesses
of normal layer at different temperatures. A hysteretic peculiarity
of $V(I)$ dependence is described as result of the negative
differential resistance. The current-voltage characteristic of
high-$T_c$ composite YBCO +BaPbO${_3}$ were computed.

\end{abstract}

\maketitle

\section{Introduction}
Superconductor -- normal metal -- superconductor (SNS) junctions
have the current-voltage characteristics (CVCs) with rich
peculiarities. Given certain parameters of junction, CVCs of SNS
junctions demonstrate the excess current, the subharmonic gap
structure and the negative differential resistance at low bias
voltage. The region of negative differential resistance corresponds
to the hysteresis of voltage in the bias current measurements. The
SNS junctions with nonlinear CVCs are promising for different
applications, e.g. low-noise mixers in submillimetre-wave region
\cite{nic3g,matsu}, switcher \cite{mamal}, nanologic circuits
\cite{hu4}.

Description of CVCs of SNS junctions was subject of many articles
and there were recognized the key role of multiple Andreev
reflections \cite{kbt,octfh,kgn,guza,brat,baav}. The main features
of CVCs enumerated above are successfully described by K\"{u}mmel -
Gunsenheimer - Nicolsky theory (KGN) \cite{kgn}. KGN theory is
applicable for the thick and clean weak links, where the normal
metal layer N has the thickness $2a$ larger than the coherence
length of superconductor, the inelastic mean free path $l$ larger
than $2a$. A simplified model in frame of KGN theory was developed
by L.A.A. Pereira and R. Nicolsky \cite{nic}. This simple model is
relevant for the weak links with thin superconducting banks S. The
contribution of scattering states \cite{kgn} is omitted in the model
\cite{nic}.

Pereira - Nicolsky and KGN model were applied earlier to describe
experimental CVCs of various weak links
\cite{nic2,PphC99,Pftt02,Pftt03,PphC04,go04}. Experience of
applications demonstrates that oversimplified Pereira - Nicolsky
model gives only qualitative description. We suggest a new simple
modification of KGN theory. It is shown that the CVCs of SNS
junctions can be computed without all the complex Ansatz of KGN
theory. We hope this will lead to more extensive using of KGN based
approach to the calculation of weak link characteristics.

\section{Current-voltage characteristics}

\subsection{Model}

\begin{figure}[htbp]
\centerline{\includegraphics[width=2.39in,height=2.26in]{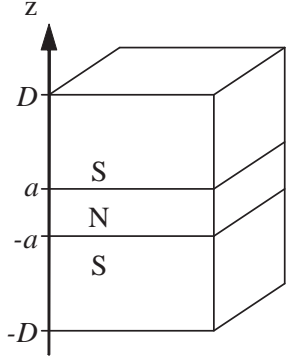}}
\caption{SNS junction.} \label{fig1}
\end{figure}

Let us consider a voltage-biased SNS junction with a constant
electric field which is in negative z direction perpendicular to the
NS interfaces and exists in the N layer only (\Fref{fig1}). The
normal layer has the thickness $2a$. The thickness of
superconducting bank is $D-a >> 2a$.

Dynamics of quasiparticles in the SNS junction was considered in the
work \cite{kgn}, where the time dependent Bogoliubov - de Gennes
equations are solved for the wave packets of nonequilibrium
electrons and holes. The expression for dissipative current density
in SNS junction was resulted in \cite{kgn}. Calculations of density
of states \cite{plegk} and probability of finding of the
quasiparticles \cite{kgn} in the N region are requested preliminary
to calculate the current. The expression for current through SNS
junction with thick superconducting banks ($D-a >> 2a$) can be
written as following \cite{kgn}:

\begin{eqnarray}
\label{eq1IV}  I(V) = \frac{e\hbar}{2am^\ast } \sum\limits_{n =
1}^\infty {\exp \left( {- \frac{2a}{l} n} \right)}  \nonumber \\
\int\limits_{ - \Delta + neV}^{\Delta + eV} dE \: \sum\limits_{r}
g_r \left( E \right) P_N \left( E \right) k_{zF} \tanh \left(
{\frac{E}{k_b T}} \right) + \frac{V}{R_N} \,,
\end{eqnarray}

\noindent where $g_r(E)$ is the two dimensional density of states,
$P_{N}(E)$ is the probability of finding of the quasiparticles with
the energy $E$ in the N region, $m^\ast$ is the effective mass of
electron, $l$ is the inelastic mean free path and $R_{N}$ is the
resistance of the N region with the thickness $2a$, $\Delta$ is the
value of energy gap of superconductor at the temperature $T$,
$k_{zF}$ is the z component of Fermi wave vector of quasiparticles,
$n$ is the number of Andreev reflections which quasiparticles
undergo before they move out of the pair potential well.

The probability $P_{N}(E)$ of finding of quasiparticles with energy
$E$ in the N region is given by Eq.(2.19) of \cite{kgn}:

\begin{equation}
\label{eq_pn} P_N(E) = \frac{2a}{2 a + 2 \lambda}
\end{equation}
\noindent with the penetration depth  $ \lambda =
\frac{\hbar^2}{m^\ast} \frac{k_{zF}}{\sqrt{\Delta ^2 - E^2}}$ for $E
< \Delta$, $\lambda < D - a$ and $\lambda = D - a$ otherwise. For
the quasiparticles from the scattering states $P_N(E)= 2a/2D$. Let
us accept for the sake of simplicity $\lambda >> a$. Therefore
$P_N(E)= 2a/2\lambda$ for the bound states.

\subsection{Density of states}

The density of states \cite{plegk} is found from

\begin{equation}
\label{eq_dos} g_r(E) = \frac{A}{\pi} \sum\limits_{r} k_{zF,r}
\left| {\frac{dE}{dk_{zF}}} \right|^{-1}_{k_{zF,r}},
\end{equation}
\noindent where $A$ is the normal layer area, $k_{zF,r}$ defines
the value of $k_{zF}$ for which $E_r = E$.

\begin{figure}[htbp]
\centerline{\includegraphics[width=87.5mm,height=63mm]{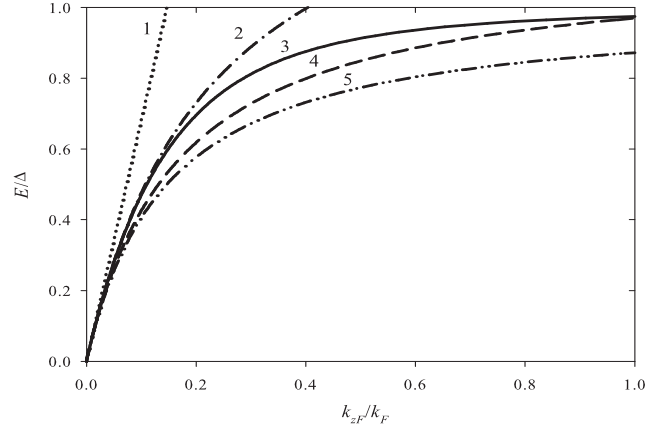}}
\caption{Energy of the bound Andreev state with $r=0$; $2a=5000$
{\AA}; $\Delta = 0.57$ meV; $k_F = 1.62$ {\AA}$^{-1}$. 1)
Eq.(\ref{eq4c}), $C=0$; 2) Eq.(\ref{eq4c}), $C=1$; 3) the exact
solution of Eq.(\ref{eq2er}); 4) Eq.(\ref{eq4c}), $C= \pi/2(1-
am^\ast\Delta/\hbar^2k_{F})$; 5) Eq.(\ref{eq4c}), $C=\pi/2$.}
\label{fig2st}
\end{figure}

The energy spectrum $E_r(k_{zF})$ consists of the spatially
quantized bound states and the quasicontinuum scattering states. The
energy eigenvalue equation for the spatially quantized bound Andreev
states \cite{kgn} is transcendental; it calculated numerically only
\Fref{fig2st}:

\begin{equation}
\label{eq2er} E_r \left( {k_{zF} } \right) = \frac{\hbar ^2k_{zF}
}{2am^\ast}\left( {r\pi + \arccos \frac{E_r }{\Delta }} \right),
\end{equation}

\noindent where $r$ = 0,1,2,\ldots

Let us simplify Eq.(\ref{eq2er}). The expansion of
arccos($E_{r}/\Delta )$ in (\ref{eq2er}) to Taylor series $(\pi/2 -
E_{r}/\Delta $ +\ldots ) up to the second term and the subsequent
expressing of $E_r(k_{zF})$ are executed. Then we inserted the
correcting multiplier $C$ for best fitting of Eq.(\ref{eq2er}).

\begin{equation}
\label{eq4c} E_r \left( {k_{zF} } \right) \approx \frac{\hbar
^2k_{zF} }{2am^\ast}\pi \left( {r + \textstyle{1 \over 2}} \right) /
\left( {1 + C\frac{\hbar ^2k_{zF} }{2am^\ast \Delta}} \right)
\end{equation}

If $C=0$ then the spectrum of Pereira - Nicolsky model is reproduced
(curve 1, \Fref{fig2st}). R. K\"{u}mmel used Eq.(\ref{eq4c}) with
$C=\pi/2$ \cite{kue_pr} for approximated calculation of the energy
spectrum (curve 5, \Fref{fig2st}).

The density of bound states follows from (\ref{eq_dos}) and
(\ref{eq4c})

\begin{equation}
\label{eq5gbs} g_{r}\left( E \right) = \frac{A}{\pi }\left(
{\frac{2m^\ast a}{\hbar ^2}} \right)^2\sum\limits_r {\frac{E}{\pi
^2\left( {r + \textstyle{1 \over 2}} \right)^2\left( {1 -
C\frac{E}{\pi \left( {r +\textstyle{1 \over 2}} \right) \Delta }}
\right)^3}}
\end{equation}

For quasiparticles from the quasicontinuum states, the energy
spectrum is approximated by the continuous BCS spectrum of
homogeneous superconductor \cite{kgn,plegk}:

\begin{equation}
\label{eq_ess} E(k_{zF}) = \sqrt{\left(\frac{\hbar^2}{2 m^\ast}
\left(k_F^2-k_{zF}^2\right)\right)^2 + {\Delta^*}^2}
\end{equation}

For SNS junction with thick superconducting banks, the effective
energy gap $\Delta^*$ equals $\Delta$. Then the density of
quasicontinuum scattering states is

\begin{equation}
\label{eq7gss} g(E) = \frac{A}{\pi^2 } \frac{2m^\ast}{\hbar ^2}
k_F D \frac{E}{\sqrt {E^2 - \Delta ^2} }
\end{equation}

\begin{figure}[htbp]
\centerline{\includegraphics[width=87.5mm,height=63mm]{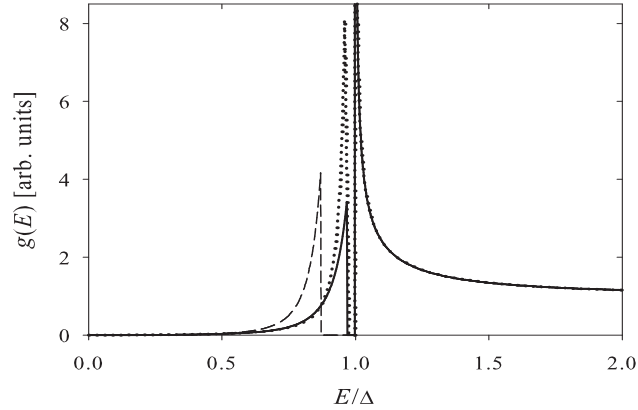}}
\caption{Density of states $g(E)$ of SNS junction with thick
superconducting banks resulted by \cite{plegk} (dotted line), $g(E)$
calculated by Eq.(\ref{eq4c}) with $C=\pi/2$ (dash line), $g(E)$
calculated by Eq.(\ref{eq4c}) with $C = \pi/2(1- am^\ast\Delta/\hbar
^2k_{F})$ (solid line). $D=70000$ {\AA}; $2a=5000$ {\AA}; $T_c=3.77$
K; $k_F = 1.62$ {\AA}$^{-1}$.} \label{fig3dos}
\end{figure}

It is reasonable to choose the variable multiplier $C$ from the best
fitting of the energy spectrum and the density of states. We suggest
$C= \pi/2(1- am^\ast\Delta/\hbar ^2k_{F})$ for $C>1$ and $C=1$
otherwise. Such choice of $C$ provides a good agreement of
Eq.(\ref{eq4c}) (curve 4, \Fref{fig2st}) with the numerical solution
of Eq.(\ref{eq2er}) for different relation of $a,m^\ast,\Delta,k_F$.
The coincidence of density of states resulted with $g(E)$ by
\cite{plegk} is satisfactory (\Fref{fig3dos}).

\subsection{Current density}

The current density of quasiparticles from bound states is resulted
with (\ref{eq1IV}) and (\ref{eq5gbs}):

\begin{eqnarray}
\label{eq_jbs} j_{bs} (V) = \frac{e{m^\ast}^2 a^2}{2 \pi^3
\hbar^5} \sum\limits_n \exp \left( { - \frac{2a}{l} n} \right)
\qquad\qquad\qquad\qquad\qquad\qquad\qquad \nonumber
\\* \int\limits_{ - \Delta + neV}^\Delta {dE\sum\limits_r
{\frac{\left| E \right|\sqrt {\Delta ^2 - E^2}}{\left( {r +
\textstyle{1 \over 2}} \right)\left( {1 - C\frac{\left| E
\right|}{\pi \Delta \left( {r +\textstyle{1 \over 2}} \right)}}
\right)^3}} \tanh \left( \frac{E}{2k_B T} \right) }
\end{eqnarray}

After substitution of (\ref{eq7gss}) to (\ref{eq1IV}) and exclusion
of small term, we get the current density of quasiparticles from
quasicontinuum states:

\begin{equation}
\label{eq9jss} j_{ss} (V) = \frac{e}{4 \pi^2 \hbar } {k_F}^2
\sum\limits_n {\exp \left( { - \frac{2a}{l} n}
\right)\int\limits_{E_1 }^{\Delta + eV} {dE\frac{E\tanh \left( {E /
2k_B T} \right)}{\sqrt {E^2 - \Delta ^2} }} },
\end{equation}

\noindent where $E_{1}$ = $-\Delta + neV$ for $-\Delta + neV \ge
\Delta $ and $E_{1}=\Delta $ otherwise.

The current densities (\ref{eq_jbs}) and (\ref{eq9jss}) include the
voltage dependence only within the integral limits.

If $eV >> k_{B}T$, $\Delta $ the integral in (\ref{eq9jss}) can be
transformed and the excess current density is resulted:

\begin{equation}
\label{eq10} j_{ex} (V) = \frac{e}{2 \pi^2 \hbar } {k_F}^2 \Delta
\tanh \left( \frac{eV}{2k_B T} \right)\exp \left( - \frac{2a}{l}
\right)
\end{equation}

This excess current density is the same as one obtained in KGN
theory (Eq.(4.12) in \cite{kgn}).

Note that $j_{bs}(V)$ dependence does not change practically if
the second summation in (\ref{eq_jbs}) is interrupted at $r = 0$.
Therefore we can write the expression for total current density as
follows

\begin{eqnarray}
\label{eq11tot} j(V) = \sum\limits_n \exp \left( { - \frac{2a}{l}
n} \right) \Biggl \lbrace \frac{2 e {m^\ast}^2 a^2}{\pi^3 \hbar^5}
\int\limits_{-\Delta + neV}^\Delta dE {\frac{\left| E \right|\sqrt
{\Delta ^2 - E^2}}{\left( 1 - C \frac{2 \left| E \right|}{\pi
\Delta} \right)^3}} \tanh{\left( \frac{E}{2k_B T} \right)}
\nonumber
\\* + \frac{e {k_F}^2 }{4 \pi^2 \hbar} \int\limits_{E_1}^{\Delta + eV}
dE \frac{E}{\sqrt {E^2 - \Delta^2}} \tanh{\left( \frac{E}{2k_B T}
\right)} \Biggr \rbrace + \frac{V}{R_{N} A} \qquad\qquad
\end{eqnarray}
$C= \pi/2(1- am^\ast\Delta/\hbar ^2k_{F})$ for $C>1$ and $C=1$
otherwise; $E_{1}$ = $-\Delta + neV$ for $-\Delta + neV \ge \Delta
$ and $E_{1}=\Delta $ otherwise.

This simplified model allows to calculate the CVCs of weak links
with thick $D-a>>2a$ superconducting banks. It operates for
different thicknesses of normal layer $2a<l$ and different
temperatures lower $T_c$.

\section{Comparison with experimental current-voltage
characteristics}

\begin{figure}[htbp]
\centerline{\includegraphics[width=87.5mm,height=63mm]{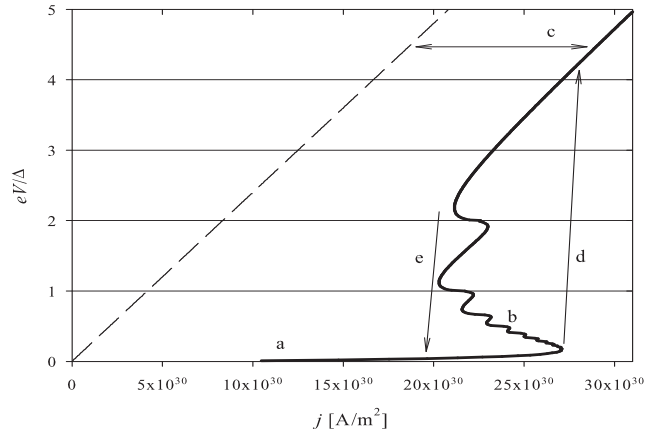}}
   \caption{Voltage-biased current-voltage characteristic of SNS junction with steep rise of current (a), region of negative differential resistance (b), excess current density $j_{ex}$ (c).
    $T_c = 3.77$ K, $\Delta_0 = 0.57$ meV, $k_F = 1.62$ {\AA}$^{-1}$, $2a = 5000$ {\AA}, $l = 15 a$, $T = 0.1 T_c$.
    Arrows (d) and (e) display the hysteretic jumps of voltage in current-biased CVC}
   \label{figmodela}
\end{figure}

The model is successfully applied for the description of CVCs of SNS
junctions \cite{arx_g06}. The steep rise of current density at low
voltage, the arches of subharmonic gap structure (SGS), and the
excess current are reproduced (\Fref{figmodela}) in the computed
CVCs (\ref{eq11tot}). The multiple Andreev reflection is the main
reason for these peculiarities \cite{kgn}. Position of the (n-1)th
arch of SGS is between $V_n$ and $V_{n+1}$, where $V_n =
2\Delta/(n-1)e$, so the largest 1th arch is between $\Delta$ and
$2\Delta$. Small peaks on the arches near $V_n$ are caused by subgap
peak on $g(E)$.

The number of allowed Andreev reflections decreases with increasing
of bias voltage \cite{kgn}. The current density due to Andreev
reflections decreases correspondingly. In that time the ohmic
current density increases as well as voltage. Region of the negative
differential resistance appears in the CVC if the decreasing of
Andreev current density is stronger than the increasing of the ohmic
current density. The measurement of current-biased CVC demonstrates
the hysteretic voltage jumps (\Fref{figmodela}) instead the region
of negative differential resistance.

To prove the model we firstly \cite{arx_g06} described the CVCs of
tin microbridges \cite{guba,oct,guba1} in different temperatures.
The hysteretic peculiarity is absent in the CVCs. Satisfactory
agreement of the calculated curves and the experimental data is
achieved \cite{arx_g06} for the known parameters of Sn.

The application of model is possible to describe the CVCs of
combination of weak links e.g. networks and contacts connected in
series. The networks of weak links, which realize in the
polycrystalline high-$T_c$ superconductors, have the composite CVCs.
These CVCs are a superposition of individual CVCs of the single weak
links that constitutes the network. Fitting parameters for current
and voltage should be used in the model to account a straining of
CVC along $I$ and $V$ axes \cite{nic,PphC99}. The weak links
connected in series (SNSNS...) can be realized in the break
junctions and wires with phase slip centers. In this case, the
straining of CVCs along $V$ axis is accounted by the formula for
series of weak links with dispersion of parameters
\cite{Pftt02,Pftt03,PphC04}.

\begin{figure}[htbp]
\centerline{\includegraphics[width=87.5mm,height=63mm]{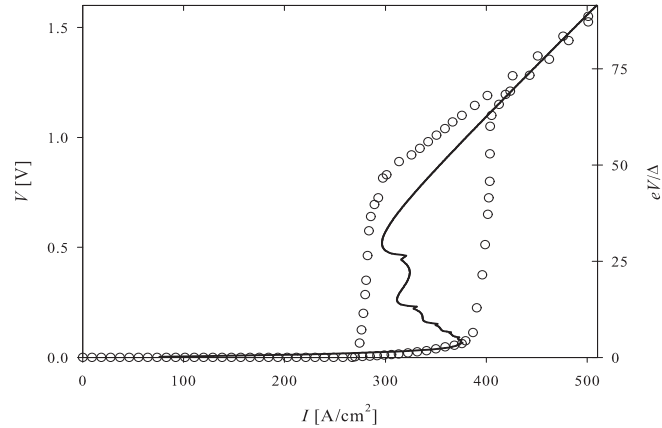}}
   \caption{Current-voltage characteristic of composite YBCO + BaPbO${_3}$.  Experiment \cite{PphC99} (points) and calculations (solid lines).}
\label{figcvcp}
\end{figure}

We used the developed model to compute the CVC of composite 92.5\%
YBCO + 7.5\% BaPbO$_3$ \cite{PphC99} that is the network of weak
links. We get the satisfactory agreement (\Fref{figcvcp}) for the
parameters of investigated composite ($T_c = 93.5$ K, $\Delta_0 =
17.5$ meV, $k_F = 0.65$ {\AA}$^{-1}$, $2a = 50$ {\AA}, $l = 9 a$,
$m^* = 4 m_e$, $T =4.2$ K).  The scale of right axis in units
$eV/\Delta$ demonstrates the straining of CVC due to the
superposition of individual CVCs of single weak links. This
description of experimental CVC is more correct than earlier one
\cite{PphC99} by Pereira-Nicolsky model.

\section{Conclusion}

The simplified model for calculation of current-voltage
characteristics of SNS junctions was developed. The KGN approach
\cite{kgn} was changed to be more convenient for description of
experimental CVCs of weak links with thick superconducting banks.
The model operates for different thicknesses of normal layer $2a<l$
and different temperatures lower $T_c$. The frequent observed
peculiarities (steep rise of current, arches of subharmonic gap
structure, negative differential resistance, excess current) in CVCs
of SNS junctions are interpreted to be produced by the multiple
Andreev reflections. The hysteretic peculiarity is described as
result of the negative differential resistance.

The model was applied to compute
the hysteretic current-voltage characteristics of high-$T_c$ composite YBCO +BaPbO${_3}$.\\

\section*{Acknowledgements}

I am thankful to D.A. Balaev, R. K\"{u}mmel and M.I. Petrov for
fruitful discussions. This work is supported by program of
President of Russian Federation for support of young scientists
(grant MK 7414.2006.2), Krasnoyarsk Regional Scientific Foundation
(grant 16G065), program of presidium of Russian academy of science
"Quantum macrophysics" 3.4, Lavrent'ev competition of young
scientist projects (project 52).

\end{document}